\documentclass[aps,twocolumn]{revtex4-1}
\usepackage{amssymb}
\usepackage{amsmath}
\usepackage{amstext}
\usepackage{amsfonts}
\usepackage{bbold}
\usepackage{slashed}

\usepackage{graphicx}

\def\be{\begin{eqnarray}}
\def\ee{\end{eqnarray}}

\def\v{v_{\rm F}}
\def\L{{\rm L}}

\begin{document}

\title{Field Theory for Zero Sound   and  Ion Acoustic Wave in Astrophysical  Matter} 

\author{Gregory Gabadadze}
\affiliation{Center for Cosmology and Particle Physics, Department of Physics,\\
  New York University, New York, NY 10003, USA}
\affiliation{NYU-ECNU Joint Institute of Physics at NYU-Shanghai, Shanghai 200062, China}
\author{Rachel A. Rosen}
\affiliation{Physics Department and Institute for Strings, Cosmology, and Astroparticle Physics,\\
  Columbia University, New York, NY 10027, USA}

\begin{abstract}
\vspace{0.2cm}

We set up  a field theory model to describe the longitudinal low energy modes in 
high density matter present in white dwarf stars. At the relevant 
scales,  ions  -- the nuclei of oxygen, carbon and helium -- 
are treated as heavy point-like  spin-0  charged particles  in an effective field theory approach, 
while  the  electron dynamics is described by the Dirac Lagrangian at the one-loop level.  
We show that  there always exists a longitudinal 
gapless mode in the system irrespective whether the ions are in a  
plasma, crystal,  or quantum liquid state. For certain values of the parameters,  the 
gapless mode can be interpreted as  a zero sound mode and, for other values,  as an ion 
acoustic wave; we show that the zero sound and ion acoustic wave are complementary to each other.
We  discuss possible physical consequences of these modes  for properties of white dwarfs.  

\end{abstract}

\maketitle

\section*{Introduction and Summary}
\vspace{-.25cm}

Consider  a neutral system of particles  made of  quantum degenerate electrons,  
and either Oxygen ($Z=8$), or Carbon ($Z=6$),  or Helium  ($Z=2$) nuclei, at 
mass densities $\rho \sim (10^6-10^7)$ g/cm$^3$, and temperatures $T< (a~few)\times 10^8~K$.  
For such high densities, average inter-particle separations in the system are  much smaller than the 
atomic scale -- hence no atoms would form,  even if the system were cold. 
On the other hand,  the separations are much larger than  the 
nuclear scale,  hence,  one can regard the $O,C$, and $He$ nuclei as point-like, 
positively charged,  spin-0 particles; we'll refer them as $\it ions$ in 
the present work, to emphasize that one  doesn't  
need to care about their detailed nuclear structure. 

For the $O,C$, and $He$ ions the baryon number, $A$, equals  twice their  
charge $Z$, and therefore the ion mass densities,    $\rho \sim (10^6-10^7)$ g/cm$^3$, 
translate into the electron $number~density$   that can be estimated as   
$J_0 \simeq  (\rho/2 m_p) \sim(0.1 -0.3\,{\rm MeV})^3$, where $m_p$ is the protons mass. 
The corresponding Fermi momenta are $p_F \sim (0.3 - 0.9){\rm MeV}$, 
and hence the electrons are (nearly) relativistic.  Their Fermi energy  significantly exceeds 
their thermal energy, as well as their two-body  Coulomb interaction energies;  
therefore, the electron system can be regarded as a quantum degenerate gas.

As  for the ions, they're much heavier than the electrons, and hence  
their thermal de Broglie wave-lengths much shorter, so that at  $T\sim  (a~few)\times 10^8~K$ 
they generically form a classical gas (see \cite {Teukolsky}, and references therein). 
However, at lower temperatures two effects need to be taken into account:
(a) Their  two body Coulomb interaction energies  start to dominate over their 
thermal  energy; (b) Their thermal de Broglie wavelengths  become comparable 
to their average separation,  and they enter  a quantum regime.  As a result, 
below $T\sim   (a~few)\times 10^6~K$, the ions may form either a classical or  quantum crystal 
\cite {MestelRuderman}, or a quantum liquid \cite {Gabadadze:2008mx}, depending on concrete values of $\rho$ and $Z$.

The above described matter  is believed to exist in the interiors of white dwarf stars (WD's). 
These are stars that  finished their thermonuclear burning process, and are sitting in the sky to radiate 
away the heat stored in them. They can be regarded as retired stars,  slowly evolving from being 
part of luminous matter to become  baryonic dark matter. There are abundant numbers of such WD's  
observed in our galaxy alone, and  the interiors of the majority of  them consist of Carbon, or Oxygen,  or 
a mixture of the two, with mass density $\rho \sim (10^6-10^7)$ g/cm$^3$, although higher densities 
are also present in many of them  (see, e.g., \cite {Teukolsky} and references therein).  A typical  
WD starts off at temperature $\sim  (a~few)\times 10^8~K$, and  takes  from   a few   to 10 Gyrs to cool 
down to about $10^5~K$ or below, becoming then directly unobservable.

The WD cooling  rate is strongly influenced by thermodynamic properties of the state 
of matter in the bulk of these stars, and in particular by  its specific heat.
The latter  can be calculated  if one knows dispersion relations of low energy 
excitations of the substance in WD's \cite {Teukolsky}.  As mentioned above, depending on concrete values 
of the temperature $T$,   density $\rho$ , and ion charge $Z$, the ions  could end up being  in a 
classical gas state,  or may create a  classical or quantum crystal,  or may 
condense into a  quantum  liquid state. These different substances  will have different  low energy excitations, 
and hence,  different specific heats and cooling rates.  Knowing accurate values of these rates is 
important,  e.g., for precise determination of the age of the Universe.

Irrespective of the microscopic structure  of the resulting state of the ions in WD's,  
the ion system may be treated as a uniform substance described by an appropriate equation 
of state,  at length scales much greater than  the average inter-ion separation. 
We'd like to study  long-wavelength collective fluctuations 
in this neutral system.  Furthermore, we'll be solely interested  in the longitudinal 
low energy excitations  for reasons that will become clear below. 
In particular, we'd like to understand the interplay between the zero sound mode, that 
typically exists in degenerate fermionic systems,  and the ion  acoustic wave that is 
usually present in a neutral plasma.  What we'll show is that these two modes are complementary to each 
other:  when one of these modes is present  the other one is absent.  However,  one of them is 
always present.  We  will also show that  the cores of $O$ and $C$ WD's with  
$\rho \sim (10^6-10^7)$ g/cm$^3$ support only the ion acoustic wave;  
the zero sound mode could be present in those WD's in a relatively narrow 
spherical shell, away from the cores if density in those domains is  $\rho < 10^5$ g/cm$^3$.

In Section 1 we will set up a prototype  effective field theory model 
to calculate dispersion relations  for  the  low energy excitations.   Our  calculations 
will focus on temperatures below $(a~few)\times 10^7~K$.  The two-body 
Coulomb energy of the ions is of the order of,  $ (10^4 -10^5) ~eV=(10^8 -10^9) ~K$.
While the Coulomb interactions are screened by the electrons, the screening length 
is greater than the average inter-ion separation, thus leading  only to a small reduction, at the level of 
10$\%$ or so,  of the unscreened two-body Coulomb interaction energy \cite {Ichimaru:1982zz}.
Therefore,   for  $T< (a~few)\times 10^7~K$,   the screened Coulomb energy  
dominates over the ion thermal energies,  by at least an order of  magnitude. 
Hence, we neglect the thermal effects. 
Likewise,  finite temperature effects are negligible  for the calculation of  dispersion 
relations in the fermionic part of the system,  since $T/J_0^{1/3}\ll 1$.  
Thus, in Section 1, we  formulate an effective field theory at zero temperature, 
to calculate  the dispersion relations  for low energy modes.   
One can  then use the standard formalism of finite temperature statistical mechanics to evaluate  the 
effects of the dispersion relations  on  the  values of the specific heat.  
This is a self-consistent procedure,  as long as $T/J_0^{1/3} \ll1$.  

In Section 2 we recover all the results of Section 1 in a Coulomb gauge.
The advantage of the latter is that the properties of the longitudinal collective modes
are  captured  by the phase of a scalar field  describing the system of 
spin-0 charged ions. It is in this section that we  show that the $O$ and $C$ 
WD's will dominantly support the ion acoustic wave in their cores, even 
before the ions turn into a crystal state.

Section 3 is dedicated to  the $He$ WD's. There are WD's  that have a helium core
due to removal of matter and energy  from them  by their binary companions (see discussions and references
in \cite {Gabadadze:2009jb}).  Among these, furthermore, there is a very small subclass of the dwarf stars for
which the temperature $T_c$ at which the de Broglie wavelengths of the nuclei begin
to overlap, is higher than the would-be crystallization temperature. 
Then, right below $T_c$, the quantum-mechanical uncertainty in the position of the charged 
nuclei is greater that the average internuclear separation. This is  exactly 
opposite to the crystallized state where the nuclei would have well-localized positions 
with slight quantum-mechanical fuzziness due to their zero-point oscillations.

It was argued in Refs. \cite {Gabadadze:2008mx,Gabadadze:2008pj} 
that such a system, instead of forming a crystalline lattice,  would  condense
owing to the quantum-mechanical probabilistic ÒattractionÓ of Bose
particles to occupy one and the same zero-momentum state, and leading to a quantum
liquid in which the charged spin-0 nuclei would form a macroscopic quantum
state with a large occupation number Ð the charged condensate.

The  dispersion relations for the quasi-modes of the charged condensate were   derived 
in \cite {Gabadadze:2008mx,Gabadadze:2008pj}. These  results were obtained by using the unitary gauge, 
and  the Thomas-Fermi (TF) approximation for the electrons.
Using these results the cooling  of the $He$ core WD stars was  studied  in 
\cite {Gabadadze:2009dz, Gabadadze:2009jb}, with the conclusion  that  they'd cool faster due to condensation, and 
that this prediction could be tested if a large enough  sample of  $He$ core WD's existed.   

However,  it was subsequently shown in \cite {Bedaque:2011hs},  the TF  approximation  combined with the 
unitary gauge,   misses  one gapless mode, which could affect the cooling calculation.  
In particular, instead of using the TF  approximation, Ref. \cite {Bedaque:2011hs}, took into account  
the  one-loop fermion effects,   and unveiled the  gapless mode.  Furthermore, it was also 
shown in \cite {Bedaque:2011hs}, that  the gapless mode  makes a negligible contribution to the specific heat 
at relevant temperatures and densities for $He$  WD's, and therefore the  predictions  for the fast
cooling rates obtained in \cite {Gabadadze:2009dz, Gabadadze:2009jb} remain unchanged.  

In the present work we'll confirm  the results of Ref. \cite {Bedaque:2011hs} on the existence of the gapless mode
in charged condensate,  by doing calculations in both 
unitary  and Coulomb gauges. The latter calculation turns out to be simpler, and does not require 
going beyond the TF approximation to establish the existence of the gapless mode (the one-loop result is 
needed though in the Coulomb gauge to calculate the  imaginary part  of this mode). Furthermore, the Coulomb gauge 
calculations lead us to the  arguments that  the  gapless mode found in \cite {Bedaque:2011hs}  
is an ion acoustic wave,  that  is present in charged condensate even when the finite 
temperature effects are ignored.

 A brief outlook is presented in the last section of the paper.

\section*{1. A Prototype Model}
\vspace{-.25cm}

Consider a high density medium composed of nuclei and degenerate electrons, 
as discussed in the previous section. Depending on the temperature and/or composition, 
the nuclei could  be in a state of  plasma,   crystal, or charged condensate.
In all the three cases Lorentz invariance is broken by the medium; for 
a crystal  the rotation and translation symmetries are also broken to their discrete subgroups. 
We consider this medium at  length scales larger than the average inter-particle  separations, 
and would like to study  long  wavelength longitudinal  collective modes. 
We use an effective field theory technique to write  down the Lagrangian that
respects symmetries of the system at hand.  In this section we discuss a prototype 
Lagrangian.  

The ions are much heavier than the electrons, and  at   low momenta 
the effects  they produce are  presumed to be  modeled  by introducing in the Maxwell  Lagrangian  
the ``electric" and ``magnetic" masses,  denoted by $m_0$ and $m_\gamma$, respectively 
(electrons and their effects will be accounted for a bit later)
\be
\label{L}
{\cal{L}}_{A} = {1\over 2} \left ( {E_j^2}  - {H^2_j} +m_0^2A_0^2-m_\gamma^2A_j^2 \right  )\,, 
\ee
where $E_j$ and $H_j, j=1,2,3$,  are the components of the electric and magnetic fields, and unless stated otherwise we  use the units $c=\hbar =1$, for which the vacuum values of the dielectric 
constant and magnetic permeability are set to unity.  

The presence of the ``electric" and ``magnetic" mass terms in (\ref {L})
captures the phase of the spin-0 ionic matter, as will be more clear below. 
This phase is what's capturing  the relevant low energy physics of longitudinal modes.
In reality,  the parameters $m_0$ and $m_\gamma$ 
would be scale dependent quantities, and should account for renormalization of the 
vacuum values of the dielectric constant  and magnetic permeability, due 
to the  quantum effects  of the ions.  However, we are interested in the low
momentum and frequency limit, where  these parameters can  be 
approximated by  constant $m_0$ and $m_\gamma$.  The scale dependent renormalization 
of these quantities due to the electrons will be explicitly included  below. 
We'll show that the outlined approximations give a 
reasonable description for  longitudinal modes in the 
plasma and crystal (Sections 1 and 2),  however,  they  break 
down  for the  charged condensate (Section 3), where the explicit 
scale dependence of the ``electric mass"  cannot be ignored.

In particular,  the  prototype model  allows  for a simple and clear  description 
of the conditions for the existence of a zero sound and ion acoustic wave in the 
degenerate plasma and crystal;   with a certain modification,   derived in Section 3, 
this  model  can also describe the collective fluctuations of the charged condensate.

The Lagrangian (\ref {L}) is not gauge invariant.   
The only physical meaning of the latter statement is  that (\ref {L}) describes more 
degrees of freedom than the Maxwell theory.  Local gauge invariance can always be restored at 
the expense of introducing new fields, and therefore, it is a redundancy of the description (although 
a convenient one).  In particular,   (\ref {L})  can always 
be regarded as a gauge-fixed  version of  a  gauge invariant Lagrangian
obtained from (\ref {L})  by the substitution,   $A_\mu \to B_\mu = A_\mu  -\partial_\mu \alpha$, 
with the invariance transformation,  $\delta A_\mu = \partial_\mu \gamma,~~ 
\delta \alpha = \gamma$, where $\gamma$ is a gauge transformation parameter, and 
$\mu =0,1,2,3$.   The phase field $\alpha$  makes  explicit the  presence of the  degree(s) of freedom 
beyond the two  transverse states that can be attributed to the gauge field  $A_\mu$, 
when a nonzero $\alpha$ is   retained. 

Conversely, (\ref {L})   can be 
regarded as the Lagrangian  in the  so-called unitary gauge, $\alpha=0$. 
We use this gauge in the present Section, while in Section 2 we restore 
back a nonzero $\alpha$, and use instead the  Coulomb gauge for  the gauge field,
$\partial^j A_j =0$. The results in the two gauges will naturally be the same,   but  the two 
derivations are different,  each having its own advantages  for understanding of 
the final results.

Irrespective of the gauge choice, the system described by (\ref {L}) contains  an extra  
longitudinal  degree of freedom, in addition to the two transverse  modes of a photon.
The longitudinal mode  can be thought  of as a collective low energy  excitation   
of  the charged  ion background.  Since we have not introduced yet the neutralizing electrons in (\ref {L}), 
the spectrum of  excitations is gapped by the   parameter  $m_\gamma$. As long as the latter  
is  smaller than the inverse inter-particle distance, all three  
gapped modes can still be meaningfully described by the  low energy 
Lagrangian.  

We now introduce  the electrons. Instead of writing an effective Largangian for them, we use 
the fundamental  description  in terms of the Dirac theory. This is justified: 
the ions are much heavier than the electrons,  and at energy scales below the ion mass
their collective longitudinal mode is  captured  into an effective Lagrangian (\ref {L}), while the  
electrons can be kept fundamental as long as they are  weakly interacting,  and as long as 
their loop effects  will in the end be taken into account (see below).   
Hence, the  total low-energy prototype Lagrangian   for the electron-ion 
system reads as follows:
\be
{\cal L} = {\cal L}_A+ {\cal L}_F,~~~~~
{\cal L}_F ={\bar \psi} (i {\tilde {\slashed D}} -m_e) \psi \,.
\label{Lf}
\ee
Here we have introduced the chemical potential for the electrons $ \mu_e$ 
via the usual prescription on the covariant derivative ${\tilde D}_0 = D_0- i \mu_e$. We 
will also package  the electric and magnetic fields in a Lorentz-invariant Maxwell form, 
with the Lorentz-breaking effects  of the nuclei  summarized by $m_0$ and $m_\gamma$; this is 
pending  additional Lorentz-violating effects to  arise due to the electrons. 
To account for the latter,  we look at the contribution of the electrons to the Lagrangian  
at the one-loop  level,  namely via the photon self-energy ${\bar \Pi}^{\mu\nu}$.  Then, the  
Lagrangian density for the fluctuations in the quadratic approximation is given by
\be
\label{L2}
{\cal L}&=& -\tfrac{1}{4} F_{\mu\nu}^2 +\tfrac{1}{2} m_0^2 A_0^2 -\tfrac{1}{2} m_\gamma^2 A_j^2 +\tfrac{1}{2}A_\mu
{ \bar \Pi}^{\mu\nu}A_\nu \,. 
\ee
Note that the inclusion  of a one-loop expression for ${\bar \Pi}^{\mu\nu}$  in the 
effective Lagrangian (\ref {L2}) to determine the dispersion relations, 
yields  the same results for  the dispersion relations,  as if they were deduced from the poles of Green's functions 
in which the one-loop  bubble diagrams  have been resummed (such a resummation is often  
referred as the random phase approximation).

It will be convenient to work in the momentum space,   and use the Fourier transform of ${\bar \Pi}^{\mu\nu}$
that we denote by ${\Pi}^{\mu\nu}$. Due to the conservation of the fermion number,  and rotation symmetry,   
${\Pi}^{\mu\nu}$ can be  expressed in terms of two functions,  $\Pi(\omega,k)$ and $\Pi^\bot(\omega,k)$, 
where $k=|\vec{k}|$, and takes the form
\be
\Pi^{\mu\nu} = 
\left( \begin{array}{cc}
\Pi & \frac{\omega k_j}{k^2} \Pi 
\vspace{1mm}\\
\frac{\omega k_i}{k^2} \Pi &~~ \frac{\omega^2 k_ik_j}{k^4} \Pi-\left(\delta_{ij}- \frac{k_i k_j}{k^2}\right)\Pi^\bot
\end{array} \right) \,.
\ee
 If we decompose the photon into transverse, longitudinal,  and time-like components, $A_j^\bot$, $A^\L$, $A_0$, the Lagrangian for the transverse modes decouples from that of the longitudinal and time-like components.  
In momentum space, we have
\be
{\cal L}^\bot = \frac{1}{2} A_j^\bot (\omega^2-k^2-\Pi^\bot-m_\gamma^2)A_j^\bot  \, ,
\ee
\be
\label{L2L}
{\cal L}_2^\L =
\frac{1}{2} \big(A_0~~A^\L \big) \cdot {\mathbb M} \cdot 
\left( \begin{array}{c}
A_0 \vspace{1mm}\\
A^\L 
\end{array} \right) \, ,
\ee
where
\be
{\mathbb M}= \left( \begin{array}{cc}
k^2\left(1+\frac{\Pi}{k^2}\right)+m_0^2 & \omega k \left(1+\frac{\Pi}{k^2}\right)  \\
\omega k \left(1+\frac{\Pi}{k^2}\right) & \omega^2\left(1+\frac{\Pi}{k^2}\right) -m_\gamma^2
 \end{array} \right) \, .
\ee

The dispersion relations of the two transverse modes are clearly given by
\be
\omega^2-k^2-\Pi^\bot(\omega,k)-m_\gamma^2=0 \, .
\label{trans11}
\ee
The dispersion relations for the remaining modes are given by the zeros of the determinant of the matrix ${\mathbb M}$,
\be
\det {\mathbb M} =  \left(m_0^2 \omega^2-m_\gamma^2 k^2\right)\left(1+\frac{\Pi(\omega,k)}{k^2}\right)-m_0^2 m_\gamma^2 =0\, . \nonumber \\
\label{longit11} 
\ee
If we were to neglect the contribution of the electrons, i.e., to 
set $\Pi(\omega,k)= \Pi^\bot(\omega,k)=0$, we would have  found  from (\ref {longit11}) 
one massive longitudinal mode
\be
\omega^2(k \rightarrow 0) = m_\gamma^2\, ,
\ee
alongside with  two massive transverse modes described  by (\ref{trans11}).
Let us see now how the  effects of the electrons, encoded in $\Pi(\omega,k)$ and $\Pi^\bot(\omega,k)$, 
modify this spectrum.  

The expression for the non-vacuum contribution to the one-loop self energy of the photon due to electrons at finite chemical potential is standard and can be found, for example, in \cite{LeBellac:1996,Kapusta:2006pm}.  
In the approximation of  small $\omega$ and $k$ as compared to the Fermi energy and momenta, and for 
$\omega \neq v_F k$,  this expression  
is given by 
\be
\label{Pi}
\Pi(\omega,k) \simeq  \frac{e^2 \mu_e^2}{\pi^2} \left(1-\frac{1}{2}\frac{\omega}{k\, \v}\ln{\left[ \frac{\omega+k\, \v}{\omega-k\, \v} \right]}\right) \,, 
\ee
where the Fermi velocity $\v$ is defined as  $\v= k_{\rm F}/m_e$.  
The usual Debye screening mass is denoted by $m_s$ 
\be
m_s^2 \equiv \Pi(\omega=0, k\rightarrow 0) =  \frac{e^2 \mu_e^2}{\pi^2} \, .
\ee
The above expression  is for relativistic electrons, while in the non-relativistic case 
one has: $m_s^2 = e^2 m_e k_F/\pi^2$.

Using equation (\ref{Pi}), we can now check for both massive and massless longitudinal 
modes.  For massive modes we take $k\rightarrow 0$ while keeping $\omega$ finite.  We find,
\be
\omega^2(k \rightarrow 0) = m_\gamma^2+\tfrac{1}{3}m_s^2 \v^2\, .
\label{add}
\ee
Thus the Debye screening mass contributes to the longitudinal mode of the massive photon.

To check for a massless pole, we set $\omega = x  \v k$, and also introduce the 
following notations:
\be
a^2 \equiv   \frac {m_\gamma^2}{\v^2 m_0^2} \equiv {v_0^2\over \v^2}, ~~~ b^2 
\equiv  \frac{m_\gamma^2}{\v^2 m_s^2} \equiv 
{v_s^2 \over \v^2}\,.
\label{ab}
\ee
We then take the $k \rightarrow 0$ limit while keeping $x$ fixed.  Then $\det {\mathbb M} = 0$ corresponds to,
\be
\label{detm}
\left(x^2-a^2 \right) \left(1-\frac{x}{2} \ln{\left[ \frac{x+1}{x-1} \right]}\right)=b^2  \, .
\ee
The right hand side  of the above expression is clearly both real and positive.  Thus, in order for a solution to this expression to exist, there must be some value of $x$, either real or complex, for which the  left hand side is 
both real and positive. We'll investigate these solutions below.

\subsection*{1.1  Zero Sound}

Consider real $x$ that is also large, $x \gg 1$. Then  equation  (\ref {detm}) takes the form
\be
(x^2-a^2) \left (  - {1\over 3x^2} - {1\over 5x^4}+...  \right ) =b^2,
\nonumber
\ee
and the solution is 
\be
x^2 \simeq \frac{5a^2 -3}{15b^2+5} \, .
\ee
Our assumption of $x \gg 1$ is valid as long as $a \gg \sqrt{1+3b^2}$.
In that  case, $ x^2\simeq a^2/(1+3b^2)$, and the dispersion relation for the massless mode 
is
\be
\omega \simeq  { v_0\, k \over \sqrt{ 1+ 3v_s^2 /\v^2 }}   \, .
\label{zsdispersion}
\ee
Thus, the  velocity  of the massless mode (both phase and group) is given by $v_0/\sqrt{1+3b^2}$.
The mode  exists when $v_0 \gg \sqrt{\v^2 + 3m_\gamma^2 / m_s^2}$, because  only in that case $x \gg1$.
Furthermore, in our system of electrons and ions,  $v_F \gg v_s$, and  (\ref {zsdispersion})  can be  
approximated  by the dispersion relation $\omega \simeq  { v_0}\,k$. 

Note that the slope of the  linear dispersion relation of the mode  (\ref {zsdispersion})  is 
larger than $v_F$  since $x \gg 1$,  and thus 
the $\omega(k)$ line  never intersects a region in which the near-the-fermi-surface 
modes live (see Fig. 1). Hence,  dumping of the  mode into electron-hole pairs is negligible. 

\begin{figure}
\centering
\includegraphics[scale=.24]{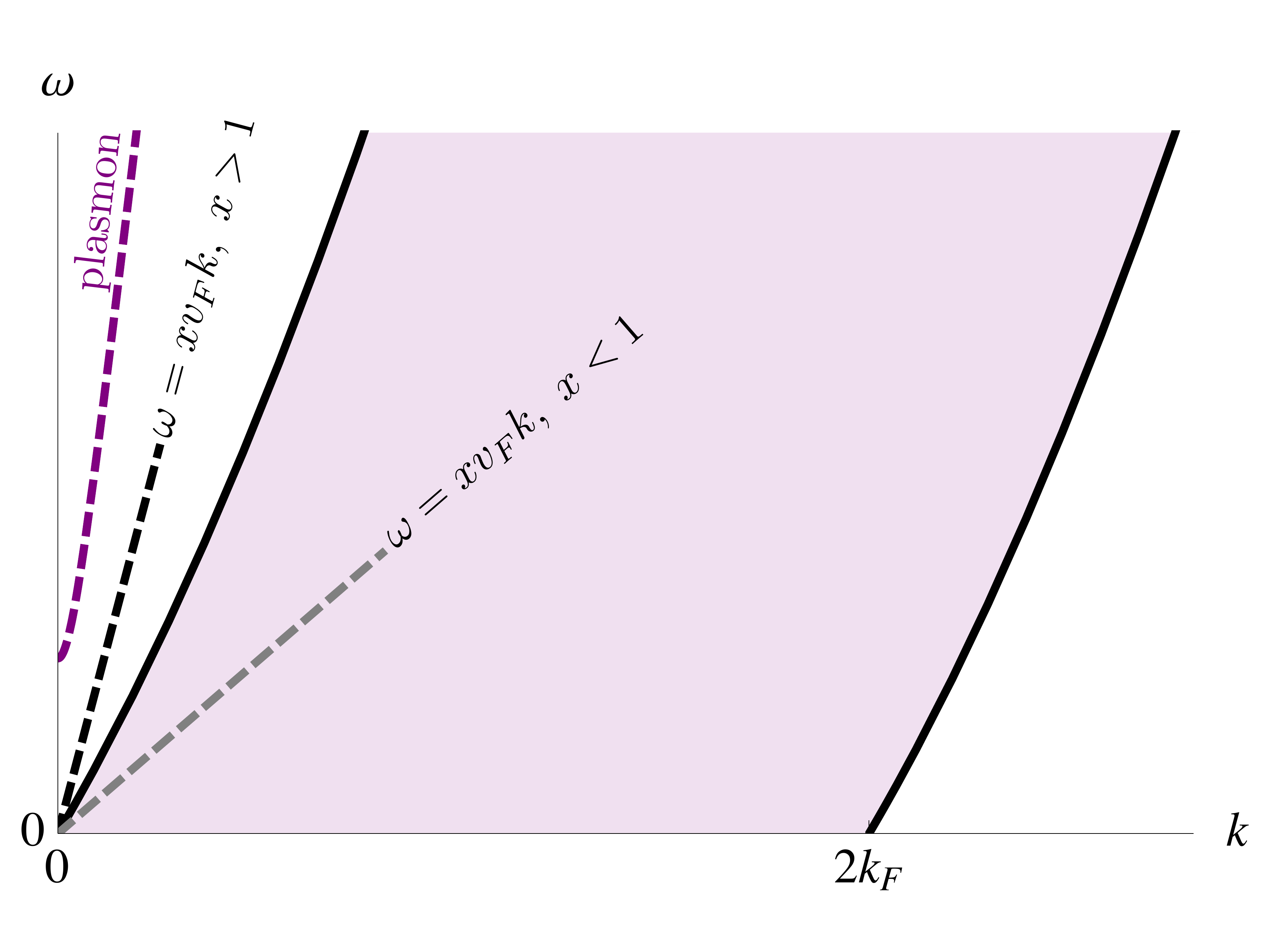}
\caption{Spectrum of longitudinal collective modes and particle-hole excitations.}
\end{figure}

This excitation  corresponds  to the zero sound mode that exists in 
interacting degenerate Fermi systems, e.g.,   in repulsively interacting Fermi liquids and gases alike 
\cite {fetterwalecka}.  More will be said on this in the next section.

\subsection*{1.2 Ion  Acoustic Wave}

Let us now turn to the  opposite regime, when $x$ could be complex but both its 
real and imaginary parts are small, i.e., $x\ll 1+i1$.  In this case, we need to take into account that  
the log entering  (\ref {detm})   is a multivalued function in the complex plane of its argument. 
Then, it is instructive to start by proving that  the equation (\ref {detm})  has no solutions on the first 
Riemann sheet   of the complex $x$-plane if $|x|<1$.

To proceed with the prove,  we introduce the definition 
\be
\label{W}
W(x) \equiv (x^2-a^2) \left(1-\frac{x}{2} \ln{\left[ \frac{x+1}{x-1} \right]}\right) \, .
\ee
Let us first consider the case that $x \in {\mathbb R}$ and $-1< x < 1$.  For this range of values we 
can use the following integral representation,
\be
W(x) =  \frac{x^2-a^2}{2} \int_{-1}^{+1}dz\,\frac{z}{z-x-i \epsilon} \, .
\ee
The epsilon prescription is given to recover the correct imaginary part of $\Pi(x)$, 
\be
{\rm Im}\, W(x) &=&  \frac{x^2-a^2}{2} \int_{-1}^{+1}dz\, z \,\pi\,\delta(z-x) \, , \nonumber \\
&=& \frac{\pi}{2} \, x(x^2-a^2)\, .
\ee
We see that for these values of $x$, $W(x)$ is complex, with the exception of $x=a$.  However, at $x=a$, we have $W(x) =0$ and thus these values of $x$ do not contain a solution of (\ref{detm}).

For all other values of $x$, real or complex, we use the following representation,
\be
\label{rep}
W(x)  =  (x^2-a^2) \int_{0}^{+1}dz\,\frac{z^2}{z^2-x^2} \, .
\ee
For both real $x$ and purely imaginary $x$, we see that $W(x)$ is real.  However, only when $a > 1$ can $W(x)$ be positive and thus solve (\ref{detm}).

Finally, we check for complex $x$.  Setting $x = \sigma+ i \beta$, with $\sigma,\beta  \in {\mathbb R}$ and $\sigma,
\beta \neq 0$, and using the representation (\ref{rep}) we find through some basic algebra that either ${\rm Im}\,W(x) \neq 0$ or ${\rm Re}\, W(x) < 0$.  Thus only in the case that $a > 1$ is $W(x)$ both real and positive.  Otherwise, we cannot solve (\ref{detm}) and no massless pole exists on the first Riemann sheet. This completes the proof. 

Thus we  look for solutions on the second Riemann sheet, and rewrite 
the equation (\ref {detm})   in the  following  form
\be
\label{detm1}
\left(x^2-a^2 \right) \left(1-\frac{x}{2}\left( \ln{\left[ \frac{1+ x}{1-x}\right]} -   i\pi \right) \right)=b^2  \, .
\ee
In the approximation  of small  $x$,  this has a solution:
\be
x \simeq \sqrt{a^2+b^2}  -  i {\pi b^2\over 4}\,.
\label{xis}
\ee
Thus the physical mode at  hand has the following dispersion relation
\be
\omega \simeq \left ( \sqrt{v_0^2 +v_s^2}  -  i {\pi v_s^2 \over 4 \v }    \right ) k \,.  
\label{disp_iaw}
\ee
The sign of the imaginary parts above  is such
that the  mode is on the second  Riemann sheet of the complex energy plane. 
Thus, it's a resonance, with a momentum dependent width.
In what follows we will argue that this is nothing but the ionic  
acoustic wave.

\vskip 0.2cm

Consider the  $m_0\to \infty$ limit;   the above mode  does exist in this limit and 
its dispersion relation reads
\be
 \omega = ( v_s  - i {\pi v_s^2 \over 4 \v }) k \, .
 \label{dispm0}
 \ee
 It is  instructive to 
go back to the Lagrangian and calculate the exact $k$ dependence of the mode, 
not only its  low $k$ limit.  The $m_0\to \infty$  limit settles us  in  the regime  $A_0=0$.
The Lagrangian then takes the form:
\be
\label{L0}
{\cal{L}} = {1\over 2} ({\partial_t A}_j)^2 -\tfrac{1}{4} F_{ij}^2 -\tfrac{1}{2}m_\gamma^2A_j^2 +
+{1\over 2} A_i {\bar \Pi}^{ij}  A_j \, .
\ee
This Lagrangian  describes four degrees of freedom:  two massive 
helicity-1, one massive  helicity-0, and one massless helicity-0  states; 
the latter  is the  mode with the dispersion relation  (\ref {dispm0}).    
To see all this we proceed by splitting the transverse and longitudinal parts 
of $A_j$  as follows:
\be
A_j = A_j^{\perp} + {\partial_j \pi  \over m_\gamma}\,.
\label{AT}
\ee
Upon this substitution the Lagrangian splits into two separate parts, one for $A^{\perp}_j$, 
\be
 {1\over 2} ({\partial_t A^{\perp}_j})^2 + \tfrac{1}{4} A_j^{\perp} \Delta A_j^{\perp}  -\tfrac{1}{2}m_\gamma^2(A^{\perp}_j)^2 
+{1\over 2} A^{\perp}_i {\bar \Pi}^{ij}  A^{\perp}_j \, ,
\nonumber
\ee
and one  for $\pi$,
\be
{1\over 2} \left ( {\partial_t \partial_j \pi \over m_\gamma } \right  )^2 - {1\over 2} (\partial_j \pi)^2 -
{(\partial_t \pi) {\bar \Pi}  (\partial_t \pi) \over m^2_\gamma}\,.  
\label{pi2}
\ee
It is straightforward to see that   the dispersion relations for the two modes in 
$A_j^{\perp}$  are determined by:
\be
\omega_{\perp}^2 = k^2 +m_\gamma^2 + \Pi^{\perp}(\omega_{\perp},k) \,,
\label{disp_T}
\ee
 while the dispersion relations that follow  from the  Lagrangian (\ref {pi2})   are determined 
by the equation  
 \be
\omega^2 = { m_\gamma^2 k^2 \over k^2 + \Pi (k, \omega)} \,.
\label{disp_pi2}
\ee
In the absence of $\Pi$ (i.e., if  the  
electrons are assumed to be frozen) we get  from (\ref {disp_pi2}) 
\be
\omega^2 = m_\gamma^2  = {Ze^2 J_0\over m_H}\,.
\ee
The latter  is a plasmon dispersion  of a charged ion gas.  In this limit of ``frozen" electrons ($\Pi\to 0$) 
there is nothing to compensate for a displaced change of ions upon their perturbation, and therefore
the mode is gapped. Note that in the $\Pi=\Pi^{\perp}=0$ limit  the longitudinal and transverse modes  
have  the same gap,  $\omega (k=0) = \omega_{\perp}(k=0) = m_\gamma$, consistent with the 
rotation symmetry.   

Once  the dynamical electrons are returned back, then the  Lagrangian 
(\ref {pi2}) and the dispersion relation  (\ref {disp_pi2})  describe two modes,  one massive and one massless.
Let us start with the massive mode, taking $k\to 0$ in (\ref {disp_pi2}) with fixed and nonzero $\omega$,
we find $\Pi \simeq - m_s^2\v^2 k^2/(3\omega^2)$, and  get a solution 
$\omega^2 = m^2_\gamma + m_s^2 \v^2/3$, which agrees with  (\ref {add}). The latter  sets  
the gap for the longitudinal photon.

For small  $\omega$  and $k$, on the other hand,  we get  a massless mode as follows:  
in this regime the real part of $\Pi$,  $Re\, \Pi \simeq m_s^2$, and therefore, 
the real part of the corresponding dispersion relation reads 
\be
\omega = \sqrt{ { m_\gamma^2 k^2 \over k^2 + m_s^2}} \simeq v_s k. 
\ee
This  dispersion relation coincides with the $m_0\to \infty$ limit of the 
one in (\ref {disp_iaw}),  up to the  imaginary part that can also  be straightforwardly 
recovered  by keeping it in $\Pi$.  Therefore, this  mode  is 
an ion acoustic wave; sometimes it's also referred as an ionic sound \cite{Dzyaloshinsky}.
It corresponds to a collective excitation of the ions  and  neutralizing fast electrons,  
producing a gapless longitudinal mode.

\vskip 0.2cm

A few comments before we turn to the next section. 
We reiterate that  the propagator in the considered regime  has no low energy poles on the first Riemann 
sheet, neither real nor   complex ones.  The obtained complex pole describing the ionic sound 
is  on the second sheet. The function that the propagator is proportional to 
$$
{1\over (x^2 -a^2) \Pi  - m_\gamma^2},
$$
has a  branch cut and exhibits a ``bump-like" behavior for real values of energy and momentum.
Why does  this  mode  have  a resonance-like nature? 
The reason is simple:  the  phase velocity of the mode  
equals to $xv_F$,  that is less than  $v_F$. The range of  allowed  energies
of the  electron-hole pairs  near the Fermi surface (depicted 
in Fig. 1)  have the upper boundary with the slope  near the origin  
that equals to $v_F$.  Thus,  a wave with the dispersion   $\omega \simeq x v_F k$ 
would be damped  due to near-the-Fermi-surface excitations, 
as long as $x<1$.  

This  also explains why this mode does not exist for $x\gg 1$; if it existed it would have 
been  faster than the near-the-Fermi-surface excitations, which  then would not be able to 
catch up with the mode to give rise  to a neutral acoustic wave;  instead, in the  $x\gg 1$ 
regime only  the zero-sound mode exists. Thus,  the zero sound and ion acoustic wave 
are complementary to each other.

\section*{2. The phase field and longitudinal modes}

The derivations  in  the previous section were performed 
in the unitary gauge, i.e.,  the phase  of the  charged scalar field 
describing collective motion of the ions was  gauge-fixed to zero. 
This was reflected in the gauge non-invariant  terms in Lagrangian (\ref {L}). 
The latter  was regarded as a gauge-fixed  version on a  gauge invariant Lagrangian,
obtained from (\ref {L})  by a substitution,  $A_\mu \to B_\mu \equiv  A_\mu -\partial_\mu \alpha$, with 
the unitary gauge corresponding  to $\alpha=0$.  In this section we'd like to  keep
nonzero $\alpha$, but we use  instead  the   Coulomb gauge 
$\partial^j A_j =0$.  This will enable us to obtain  and understand the  results of the previous 
section in a more clear way.

\subsection*{2.1 A Warmup Example}

We  start with a field theory containing a complex   scalar field,  degenerate fermions, 
and an Abelian gauge field  that couples to both the scalar and fermions  in a conventional way.
The charged scalar field  is thought to  model properties of charged nuclei, while fermions model the 
electrons,   and  the Abelian gauge boson models a photon.   
For illustrative purposes, we consider the case when the $U(1)$ gauge symmetry is spontaneously broken, 
and the gauge field acquires  the mass term as a result of this breaking. Furthermore, we consider the  
parameter space   for which the scalar field radial mode is  a heavy state  that  can be decoupled from  the 
rest of the fields.  In this approximation,  and in the Coulomb gauge, the relevant part of the Lagrangian reads 
as follows:
\be
{1\over 2} m^2 B_\mu^2 + {1\over 2} A_0 \left ( -\Delta + {\bar \Pi} \right )A_0\,.
\label{BA}
\ee
The first  term is the gauge boson mass term, while the second term 
contains a part of the Maxwell Lagrangian,  as well as the term  generated 
due to one-loop renormalization of the polarization operator via 
the fermion-antifermion pair. These are all the terms in the Lagrangian 
that contain $A_0$ and $\alpha$. 

Integrating  out $A_0$  we get for $m^2\neq 0$
\be
A_0 =  {m^2 \over \left ( -\Delta +m^2+  {\bar \Pi} \right )} \partial_0 \alpha \,.
\label{A0}
\ee
(we neglect an irrelevant  zero mode  of the inverse of the operator multiplying 
$\partial_0 \alpha$ on the right hand side of (\ref {A0})). Substituting  (\ref {A0}) 
back into the Lagrangian (\ref {BA}), and calculating the dispersion relation for 
the remaining field $\alpha$,   we obtain the following expression 
\be
\omega^2 { k^2 +\Pi (\omega, k)\over k^2 +\Pi(\omega, k) + m^2} = k^2 \,.
\label{dispAH}
\ee
Let us now study this relation in two different regimes, $x\gg 1$, and $x\ll1$, where
as before we  introduce $\omega = xv_F k$.    For the $x\gg 1$ case,  
one gets $ \Pi \simeq - m_s^2 /(3x^2)$.  Hence, the  solution to (\ref {dispAH}) reads:
\be
x^2 = {1\over v_F^2 + {3m^2/m_s^2}}.
\label{x}
\ee
When $m_s \gg m$ and $v_F \ll 1$,  we get the condition, $x\gg1$, required by the approximation made.
As per the arguments of the previous section,   
this is the dispersion relation of the zero  sound mode 
\be
\omega = {v_F \over  \sqrt{v_F^2 + {3m^2/m_s^2}} }k\,.
\ee
We see that in the Coulomb gauge the zero sound mode is described by the phase $\alpha$.  
Since $x\gg1$, the phase/group  velocity of this mode is greater 
than $v_F$, hence,  this mode experiences no damping in the approximation we use.
 
In the opposite regime, $x\ll1$, the gapless mode does not exist, as shown by {\it reductio ad absurdum}:  
assuming that $x \ll 1$, one gets 
$ \Pi \simeq m_s^2 (1 +{\cal O}(x))$, and  the solution $x= \v^{-1}\sqrt{ (1 +m^2/m_s^2)} >1$, 
which  contradicts the initial assumption that $x$ were small. 

The lack of the ion sound wave in this model has 
a reason: we used the approximation when the Lorentz-invariant vacuum condensate of the 
charge scalar  is the only source  for the mass term  (the first term 
in (\ref {BA}));  this  implies that the  number of dynamical scalars that can fluctuate is zero, 
in the  approximation used (the Lorentz invariant Higgs vacuum has zero scalar number). Hence,   
one should not expect  to have the ion acoustic  wave,  
in the approximation when the number of ions that can fluctuate is ignored
(the   charge that neutralizes the electrons in this approximation is 
not dynamical).

\subsection*{2.2  A More Realistic Model}

We now consider  the case when the ``electric" and ``magnetic" masses of the photon are different, and 
the phase is explicitly kept, while the Coulomb gauge is assumed  for $A_j$, as in the previous Subsection.
The relevant part  of the Lagrangian in this gauge then reads:
\be
{1\over 2} \left ( m_0^2 B_0^2 - m_\gamma^2 B^2_j \right )  
+ {1\over 2} A_0 \left ( -\Delta + {\bar \Pi} \right )A_0\,.
\label{BA0j1}
\ee
As before, we integrate out $A_0$ to get
\be
A_0 =  {m_0^2 \over \left ( -\Delta +m_0^2+  {\bar \Pi} \right )} \partial_0 \alpha \,.
\label{A00}
\ee
Substituting this back into (\ref {BA0j1}), and deducing the dispersion 
relation for  $\alpha$ we obtain
\be
\omega^2 { k^2 +\Pi (\omega, k)\over k^2 +\Pi(\omega, k) + m_0^2} = v_0^2 k^2, ~~~~v_0^2 
\equiv {m_\gamma^2 \over m_0^2} \,.
\label{dispAH01}
\ee
In a realistic  system such  as a plasma  and solid,  $m_0 \gg m_\gamma$, and  the 
ion sound speed, $v_0 \ll 1$. 

For $x\gg1$,  we know that $ \Pi \simeq - m_s^2 /(3x^2)$, and 
in this approximation the  solution to (\ref {dispAH01}) reads
\be
x^2 = {1\over v_F^2/v_0^2 + {3m_0^2/m_s^2}}.
\label{x}
\ee
Hence, when $m_s>m_0$ and $v_0\gg v_F$, 
we get  $x\gg 1$, and the zero sound  dispersion relation 
\be
\omega \simeq  {v_F \over  \sqrt{v_F^2/v_0^2 + {3 m_0^2/m_s^2}} }k\,.
\ee
This coincides  with the dispersion relations  for the zero sound  mode
found in (\ref {zsdispersion}). 

Unlike in the previous Subsection, however, a  solution also exists  in the opposite limit, $x\ll 1$: 
indeed, assuming  that $x\ll 1$, one gets  $ \Pi \simeq m_s^2 (1 +{\cal O}(x))$, 
and hence the solution $x={v_0\over v_F} \sqrt{ (1 +m_0^2/m_s^2)}$, 
which if,  $ {v_0} \sqrt{ (1 +m_0^2/m_s^2)} \ll v_F$ ,   
gives rise to the following physical mode
\be
\omega \simeq v_0 \sqrt{ 1 +{ m_0^2 \over m_s^2}}\, k\,.
\label{ionicsound}
\ee
This  is nothing but the dispersion relation for  
the longitudinal  ion  sound wave discussed in 
Section 1.  

One can give  heuristic arguments for the complementarity of the two modes that we've derived above. 
For this, we consider  the regime  in which $v_s$ can be neglected as 
compared with $v_0$ and $v_F$. This is a meaningful approximation  for a plasma and crystal since $m_\gamma$ 
is suppressed by the heavy ion mass scale, while $m_s$ is determined by the electron mass 
and chemical potential.  Then, the  phase velocity of both the zero sound mode and the ion acoustic wave is approximately $v_0$.  When $v_0 \ll v_F$, we  get $x \ll 1$ and the ionic sound is present; this makes sense -- 
the electrons are faster than the  wave  and thus they can readily  follow a perturbation of the 
ions to  screen  its charge and create  a gapless  neutral mode, the ionic sound.  In the opposite limit,
$v_0 \gg v_F$, the electrons are slower than the would  be ionic sound wave, therefore they cannot 
be effectively screening ion  perturbations, and it makes sense that 
the ion acoustic wave does not exist.  Instead, the electron fluctuations themselves -- which are
now effectively screened by the ambient mobile ion charge distribution --  form a collective mode, the zero sound.

Having the dispersions relations derived, let us apply them to the system of the Oxygen and Carbon 
ions at densities $\rho \sim (10^6-10^7)~{\rm g/cm^3}$.   As we've already discussed 
the corresponding Fermi momentum for the electrons is $p_F\sim (0.3 -0.9)~{\rm MeV}$. Therefore, 
the electrons are (nearly) relativistic,  with   $v_F \sim 1$,  to a good accuracy.  This implies  that the  value of 
$x$ given in (\ref {x}) can never be  greater than the unity, since $v_0<1$. Therefore, 
we conclude that the zero sound cannot be supported  in the cores of the $O$ and $C$ WD's.
Instead, the dominant longitudinal mode in this case is the ion acoustic wave (\ref {ionicsound}),
with the sounds speed approximated by $v_0 \ll 1$. The exact value of the sound speed cannot be calculated 
in our effective Lagrangian approach, however, our finding confirms the suggestion made in 
\cite {MestelRuderman}  that a longitudinal wave,  the one that's similar to the longitudinal acoustic wave 
of a crystal,  can be used to study  the cooling of  the $O$ and $C$ WD's even when the interior is in 
a strongly  interacting plasma state, and the crystal is not yet formed (i.e., from temperatures  
$(a~few)\times 10^7~K$, down to $(a~few)\times 10^6~K$,  after which the crystal forms).

As to the zero sound mode, its existence requires lower densities. While WD's  have 
fairly uniform  density profiles  in their  bulks (excluding their ``atmospheres" that are dominantly 
made of $H$ and $He$,  with some small fractions  of ``metals"), nevertheless the interiors are 
not exactly uniform. Typically one can have  variation of density amounting  to a factor of 5 
in the ratio  of   the maximal density to average  density.  Therefore,  it  might not be 
foolish  to think  of relatively narrow  spherical shells,  away from the cores of low density WD's, 
in which densities might  be $\rho < 10^5$ g/cm$^3$. In such shells the zero sound mode 
could be supported instead of the ion acoustic wave.  It could also be interesting to look  
for  the existence of the zero sound mode in the low mass brown dwarfs \cite {Berezhiani:2010db}, 
where densities are $\rho \sim 10^3~{\rm g/cm^3}$.


\vspace{.5cm}

\section*{3. Charged Condensate}
\vspace{-.25cm}
We now consider in detail the spectrum of the charged condensate.  We incorporate the effect of the background 
density of electrons through their one-loop contribution to the photon self-energy.    An analogous approach was used in \cite{Dolgov:2008pe,Dolgov:2009yt} to determine the electrostatic potential of this system at finite temperature.  More recently, this approach was applied in \cite{Bedaque:2011hs} to argue that the dynamical electrons give rise to a previously unnoticed massless mode.  Here we confirm this finding, and  argue that the obtained massless
mode is the  ion acoustic wave discussed in the previous section.

A classical nonzero vacuum expectation value of a field $\Phi$ can serve as an order parameter for the condensation of the helium-4 nuclei, describing a state with a large occupation number.  Fluctuations of the order parameter describe the collective modes of the condensate.  In the non-relativistic approximation, the effective Lagrangian for the nuclei $\Phi$ and electrons $\psi$, can be written as follows \cite{Gabadadze:2008pj}:
\be
\label{lagr}
{\cal{L}} ={\cal{P}}(X) -\tfrac{1}{4} F_{\mu\nu}^2 +{\bar \psi}(i{\slashed{D}} -m_e)\psi \, ,
\ee
where
\be
X=\frac{i}{2}\left(\Phi^*D_0\Phi-(D_0\Phi)^*\Phi\right)-\frac{|D_j\Phi|^2}{2\,m_H} \, .
\ee
The covariant derivative of the scalar field is given by $D_\mu = \partial_\mu-2ieA_\mu$.  ${\cal P}(X)$ stands for a general polynomial function of its argument. The coefficients of this polynomial are dimensionful numbers that are inversely proportional to powers of a short-distance cutoff of the effective field theory.  We normalize the first coefficient to one so that
\be
{\cal P}(X)=X+c_2 X^2+ \ldots \, .
\ee
The Lagrangian is invariant under global $U_s(1)$ transformations, responsible for the conservation of the number of scalars.  Another global $U_e(1)$ guarantees the electron number conservation.  Accordingly, we can introduce chemical potentials for both the scalars and electrons, $\mu_s$ and $ \mu_e$ respectively, via the usual prescription on the covariant derivative $D_0 \rightarrow D_0- i \mu$.

We could also have included a quartic interaction for the scalars $\lambda (\Phi^*\Phi)^2$.  However, as long as the quartic coupling is small, $\lambda\, n \ll m_H^3$, our results will not be affected by this term.

It is useful to represent the scalar as $\Phi = \Sigma\, e^{i \Gamma}$, and to work first 
in the unitary gauge where the phase of the scalar is set to zero, $\Gamma=0$.  
At a later stage, we'll move instead to the Coulomb gauge where things will become  easier.

When the scalar chemical potential is zero, $\mu_s =0$, there is a solution to the 
equations of motion of (\ref{lagr}) with a nonzero expectation value for the scalar field
\be
\langle \Sigma \rangle = \sqrt{\frac{J_0}{2}} \,,
\ee
where $J_0$ is the background electron density.

Let us consider the quadratic action around this background solution.  We introduce perturbations of the scalar field as follows:
\be
\Sigma(x) =  \sqrt{\frac{J_0}{2}} +\sqrt{m_H} \, \tau(x)\,.
\label{pert}
\ee
The factor of $\sqrt{m_H}$ has been introduced for convenience.  Let us also integrate out the electrons as was done in \cite{Bedaque:2011hs}.  Thus the Lagrangian density for the fluctuations in the quadratic approximation reads
\be
\label{lagr2}
{\cal L}_{2}&=& -\tfrac{1}{4} F_{\mu\nu}^2 +\tfrac{1}{2} m_0^2 A_0^2 -\tfrac{1}{2} m_\gamma^2 A_j^2 +\tfrac{1}{2}A_\mu\Pi^{\mu\nu}A_\nu \nonumber \\
&&- \tfrac{1}{2} (\partial_j \tau)^2 + 2m_Hm_\gamma \,A_0 \tau \,. 
\ee
Here 
\be
m_\gamma^2 \equiv (2e)^2 \, \frac{J_0}{2m_H}\,,~~m_0^2 =c_2 J_0 m_H m_\gamma^2 \, ,
\label{photonmass}
\ee
When the coefficient $c_2$ is such that $m_0 = m_\gamma$, the dispersion relations coincide 
with those of a relativistic theory, however, in general $m_0\neq m_\gamma$.

\subsection*{3.1 Spectrum in the unitary gauge}

We now consider the spectrum of these fluctuations.   We decompose the photon into transverse, longitudinal and time-like components.  In addition, let us also integrate out the scalar mode $\tau$ so that we can easily compare our results to those of the previous section.  Again, the transverse modes of the photon decouple entirely.  Their dispersion relations are given by
\be
\omega^2-k^2-\Pi^\bot(\omega,k)-m_\gamma^2=0 \, .
\ee
For the remaining modes, we have
\be
{\cal L}_2^\L =
\frac{1}{2} \big(A_0~~A^\L \big) \cdot {\mathbb M} \cdot 
\left( \begin{array}{c}
A_0 \vspace{1mm}\\
A^\L 
\end{array} \right) \, ,
\ee
where now 
\be
\label{M}
{\mathbb M} \equiv \left( \begin{array}{cc}
k^2\left(1+\frac{\Pi}{k^2}\right)+\frac{4 M^4}{k^2}+m_0^2 & \omega k \left(1+\frac{\Pi}{k^2}\right)  \vspace{0.1cm} \\
\omega k \left(1+\frac{\Pi}{k^2}\right) & \omega^2\left(1+\frac{\Pi}{k^2}\right) -m_\gamma^2 
 \end{array} \right) ,   \nonumber \\
\ee
with  $M \equiv \sqrt{m_\gamma m_H}$.  The matrix ${\mathbb M}$ differs from the analogous matrix in 
our prototype model only via the appearance of the $4M^4/k^2$ term.

The dispersion relations of these modes are again given by the zeros of the determinant of the matrix ${\mathbb M}$.    To check for the presence of massive modes, we use expression (\ref{Pi}) in the matrix (\ref{M}) and take the $k \rightarrow 0$ limit,  keeping $\omega$ finite.  Setting $\det {\mathbb M} =0$, we find a massive mode 
\be
\omega^2(k \rightarrow 0) = m_\gamma^2+\tfrac{1}{3}m_s^2 \v^2\, .
\ee
This  can be though as  the mass of the longitudinal component of the photon, that now carries  
three degrees of freedom. The mass receives two contributions, one from the charged nuclei, 
$m_\gamma^2$, and another one from  charged electrons, $m_s^2\v^2 /3$.  Hence, we find that 
the one-loop contribution shifts up the mass of the longitudinal mode. This mode 
was argued   to be heavy to contribute to specific heat of the charged condensate 
at relevant temperatures,  \cite{Gabadadze:2008pj}; we see that the one loop 
contribution makes it even  heavier, justifying  further that it can be neglected.   

To determine the possibility of a massless pole, 
we set $\omega = x v_F k$ and then take the $k \rightarrow 0$ limit.  
Then $\det {\mathbb M} = 0$ corresponds to,
\be
\label{detmtwo}
x^2 \left(1-\frac{x}{2} \ln{\left[ \frac{x+1}{x-1} \right]}\right)=\frac{m_\gamma^2}{m_s^2}  \, .
\ee
Thus we see that the charged condensate corresponds to our prototype model  of Section 1 
in the particular case with $a = 0$.  As argued above, there is a massless resonance 
solution, with the dispersion relation $  \omega \simeq  ( v_s  - i {\pi v_s^2 \over 4 \v }) k$, 
in agreement with \cite{Bedaque:2011hs}. This is also in agreement  with the dispersion 
relation  obtained  in Section 1, describing the ion acoustic wave 
in a plasma.  In our view,  this mode  has the same physical origin and interpretation 
as  the ion  sound wave of ordinary degenerate plasma \cite {Dzyaloshinsky}. 
We further strengthen this latter point in Subsection 3.3  by unveiling the 
hydrodynamics origin of the ion acoustic wave in charged condensate.

\subsection*{3.2 Spectrum in the Coulomb gauge}

Here, we derive the results of the previous subsection in the Coulomb gauge.
This makes the connection to the calculations in the prototype model presented in 
Section 2 clearer, and  helps to explain the origin of the ion sound wave 
in the charged condensate. 

As before,   we use $B_\mu = A_\mu - \partial_\mu \alpha$. The relevant part of the 
Lagrangian in the Coulomb gauge reads as follows:
\be
{1\over 2} \left ( {\bar m}_0
B_0^2 - m_\gamma^2 B^2_j \right )  + {1\over 2} A_0 \left ( -\Delta + {\bar \Pi} \right )A_0\, ,
\label{BA0j}
\ee
where the key difference from  the model of Sections 1 and 2 is that ``electric" mass
in the charged condensate  has essential dependence on the momentum:
\be
{\bar m}_0= \left (m_0^2  + {4M^4 \over  -\Delta} \right )\,.
\label{barm0}
\ee
This suggests that for small momenta, we always end up in the regime 
of large ``electric" mass for the photon; as discussed in Section 1, this 
implies that the zero sound mode will be absent, but the  ionic 
acoustic wave should be present, in agreement with the results of Subsection 3.1.
Let us see this explicitly in the Coulomb gauge:

We first integrate out $A_0$
\be
A_0 =  {m_0^2 \over \left ( -\Delta +m_0^2  + {4M^4 \over  -\Delta}  +  {\bar \Pi} \right )} \partial_0 \alpha\,.
\label{A00}
\ee
Substituting this back into the Lagrangian (\ref {BA0j}), and deducing the dispersion relation for 
$\alpha$ we get
\be
\omega^2 { k^2 +\Pi (\omega, k)\over k^2 +\Pi(\omega, k) + m_0^2 + {4M^4 \over  k^2} } 
= v_0^2 {m_0^2 \over  m_0^2 + {4M^4 \over  k^2} } k^2\,.
\label{dispAH0}
\ee
As before,  we use  the notation  $\omega = xv_F k$.  For both $k$ and $\omega$ approaching zero, we only get 
one dispersion relation, and that is with, $x\ll 1$.  This  dispersion relation  reads: 
\be
\omega \simeq {m_\gamma \over m_s} \, k\,.
\label{ionicsoundCC}
\ee
This  is the ionic  acoustic wave, in agreement with previously obtained results.  

A final comment before we move to the hydrodynamics considerations. 
In the approach  adopted above the fermions were treated  in a one loop approximation, while 
the scalars in terms of a low energy effective field theory. The quantum effects 
of the condensed scalars were captured by the order parameter Lagrangian.  
It is instructive to check   that the same results are obtained if the scalars 
are  also treated  via the one-loop calculations, as was  done in \cite{Dolgov:2008pe,Dolgov:2009yt}.
We briefly use this method  in the  Coulomb gauge. Then,  
the poles  of the full propagator are determined by 
\be
k^2 + \Pi^B(\omega, k)  +\Pi (\omega, k) =0\,,
\label{poleloops}
\ee
where the polarization operator for the  bosons, $\Pi^B(\omega, k) $ can be calculated 
straightforwardly via the corresponding  one loop diagrams \cite {Dolgov:2008pe,Dolgov:2009yt}. The part  of the loops 
that is due to the existence of the condensate reads as follows: 
\be
\Pi^B(\omega, k) = {m_\gamma^2 \over 2} [  {(2m_H- \omega )^2\over   (\omega- m_H)^2 -k^2 - m_H^2 }
 \nonumber \\ +
{(2m_H+ \omega )^2\over   (\omega+m_H)^2 -k^2 - m_H^2 }    -2      ]\,.
\label{loopPiB}
\ee
Substituting this into (\ref {poleloops}) and taking  the small momentum limit, we get 
the pole at  
\be
\omega \simeq  {m_\gamma \over  m_s}  \, k\,.
\label{ionicsoundCCloop}
\ee
The latter  coincides with  the result already obtained  above (\ref {ionicsoundCC}).

\subsection*{3.3 Hydrodynamic  considerations}

The purpose of this subsection is  to demonstrate that the ionic sound found in 
the prototype model, as well as  its counterpart emerging in charged condensate,
can be understood in terms of  standard hydrodynamics, with the only  difference  
that the  charged condensate  hydrodynamics equations  need  
to  retain  the pressure gradient term even when finite temperature effects are 
ignored, as will be shown below.    

We're looking at  a degenerate plasma of electrons and positively charged nuclei (or ions);
as already mentioned, depending on temperature $T$, the ion mass $m_H$ and charge $Z$, 
the system   of nuclei  could be in a classical gas state,  or may create a
Wigner crystal, or may be in a condensed quantum liquid state.   
In either case, at  length scales   much greater than  the inter-ion separation, 
the ion system may be treated as a uniform substance 
described by an appropriate equation of state. We'd like to understand the spectrum 
of long wavelength longitudinal collective fluctuations 
in this system.  
 
 Let us first ignore the temperature effects and consider  the case 
 when the ions are in the plasma or crystal state.  All the hydrodynamic 
 equations  presented  below  for this case are well know, but are given to 
 emphasize the difference  of these states from the charged condensate, 
 to which we'll turn  by the end of the  section.  
 
 The continuity equations for 
 the electron and  ion number densities -- denoted  respectively by $n_e$ and $n_H$   -- 
 read as:
\be
\partial_t n_{e,H} + \partial_j (n_{e,H} v_j) =0,
\label{coneq}
\ee
while the momentum equation for  the ions is
\be
\partial_t v_j + (v_k \partial_k) v_j = - {Ze \over m_H} E_j.
\label{momeq}
\ee
Consider small localized perturbations, small over-densities  $\delta n_{e,H}$  of the electrons and ions over 
their background values set by $J_0$ and $J_0/Z$, respectively
\be
n_{e} = J_0 + \delta n_{e}, ~~~n_{H} = {1\over Z} J_0 + \delta n_{H}\,.
\label{npert}
\ee
For both the plasma and crystal state at long wavelength, the Poisson equation for the electrostatic 
potential created by these perturbations reads as follows
\be
\Delta A_0 = -Z e \delta n_{H} + e \delta n_{e}\,.
\label{poispert}
\ee
For simplicity we consider relativistic electrons here.  The electron over-density  can be related to the  local potential  via  the Thomas-Fermi  
approximation:
\be
E_F \equiv \mu = -e A_0 +  p_F,
\label{EF}
\ee
and using that $n_e = p_F^3/3\pi^2$, we  find
\be
\delta n_e \simeq 3 \mu^2 e A_0 \,.
\label{epert}
\ee
Substituting  this expression into  (\ref {poispert}), one  gets: 
\be
(\Delta - m_s^2) A_0 = - {Z e} \delta n_H,
\label{Poisson1}
\ee
where $m_s^2 \equiv e^2 \mu^2 /\pi^2$ is the Debye screening 
mass squared, due to the electrons. We now look at the system of three 
equations (\ref {coneq}), (\ref {momeq}) and (\ref {Poisson1}),  
and consider their linearization above the background,  ({\it bg}),  with 
\be
n^{bg}_e=J_0,~n^{bg}_H={J_0\over Z},~A^{bg}_0 = 0,~
v^{bg}_j =0.
\label{background}
\ee
In the linearized equations we transform to the Fourier moels for  all 
perturbations, as for instance, 
\be
\delta n_e  (x,t) =\int d^3k \,d \omega \,  {\tilde{ \delta n}_e}(k,\omega)\, e^{i(\omega t- k_jx_j)}\,.
\label{Fourier} 
\ee
As a result  get the following dispersion  relation from the linearized  system of equations 
(\ref {coneq}), (\ref {momeq})  and (\ref {Poisson1}):
\be
\omega^2 = {m_\gamma^2   k^2 \over k^2 +m_s^2}, ~~{\rm where}~~ 
m_\gamma^2 \equiv {(Ze)^2 (J_0/Z) \over m_H}.
\label{ISO}
\ee
This describes a gapless collective mode; at  small momentum the dispersion relations 
reads
\be
\omega \simeq {m_\gamma \over m_s} k, 
\label{linear}
\ee
that is the dispersion relations of the ion  acoustic wave (ionic sound) for plasma, 
while  for a crystal it describes  a longitudinal acoustic phonon.

\vskip 0.2cm

Things are a bit different, however, for the case of charged condensate. 
While the continuity and Poisson  equations remain unchanged, 
the  momentum equation gets modified  due to the pressure term on the 
right hand side; this  can be shown from the Lagrangian formulation of the 
charged condensate given in Section 3 (the pressure term would certainly exist 
in ordinary plasma as a finite temperature effect, but in charge condensate it 
is  nonzero even when the finite temperature effects are ignored.)  
The corresponding momentum equation reads:
\be
\partial_t v_j + (v_k \partial_k) v_j = - \partial_j b_0 - {Ze \over m_H} E_j.
\label{momeqCC}
\ee
The difference is due to the fact  that a gradient of the pressure  is not negligible 
for perturbations in the charged condensate; the respective term is  kept as the first 
term on the right hand side in (\ref {momeqCC}); it is determined by the  gradient 
of the gauge invariant potential, $b_0 = (m_\gamma^2  - 4M^2/\Delta)\delta n_H $. 

We now combine  this new momentum equation  (\ref {momeqCC}),   with the Poisson (\ref {Poisson1})  
and continuity  (\ref {coneq}) equations, and easily derive the modified  dispersion relation for the 
longitudinal mode in charged condensate:
\be
\omega^2 \simeq {k^2 \over 4 m_H^2} +  {m_\gamma^2   k^2 \over k^2 +m_s^2}\,.
\label{ISO_CC}
\ee
The result coincides with the real part of the dispersion relation  for the massless mode
found in \cite{Bedaque:2011hs}. It also coincides, in the low momentum approximation,  
with the dispersion relations for the ion  acoustic  waves found  in Sections 1, 2, and 3  
of the present work.

An important  approximation made above is the Thomas-Fermi method. 
This  method will in general miss  fermion dynamics near the Fermi surface.  
In particular, as we see, it  does not capture  the width of the ion acoustic wave 
due to its dumping by the fermionic quasiparticles. However, this effect was already taken into account
by the one-loop consideration in \cite{Bedaque:2011hs},  and above in the present work. 

\vskip 1cm

\section*{4. Outlook}

White dwarf stars constitute  interesting physical  objects, and are also important 
for inferring key astrophysical and cosmological data.  The theory of cooling
of $O$ and $C$ WD's is well known \cite {MestelRuderman},  and agrees well with 
observations \cite {DAntona:1990ea,Hansen:2003bc}.   While at temperatures above $\sim 10^7~K$ the ions can be regarded as being  
in a classical gas state, and below $\sim 10^6$  being in  a  bcc crystal state,  
between these two temperatures -- from 
$\sim  10^7~K$, down to $ \sim  10^6~K$ -- 
one is dealing with a strongly interacting plasma of ions 
that is not easy to study using fundamental electromagnetic interactions. 
Instead, we used an effective Lagrangian approach 
that is in general well suited  to  study long-wavelength excitations, even for strongly interacting systems.    
Our finding of the absence of the zero sound mode,  and the presence of  the  longitudinal  ion 
acoustic wave in the interacting plasma regime, confirms a 
suggestion made in  \cite {MestelRuderman}  that the longitudinal acoustic wave can be  
used to describe physics of WD's in this interval of temperatures where neither gas not crystal descriptions 
are valid.

As to the zero sound mode,  its existence requires somewhat lower densities, $\rho < 10^5~{\rm g/cm^3}$. 
While WD's  have  fairly uniform  density profiles  in their  interiors, they're not exactly uniform  and are 
described  by an adiabatic equation of state. Thus,  the density can vary from the core to 
the outskirts of the bulk by a factor of 5 or more. Therefore,  there might exist   a subclass  of low density 
WD's, with  relatively narrow  spherical shells  away from the  cores,  in which densities might  
be $\rho < 10^5$ g/cm$^3$; if so then the zero sound mode could be supported in those domains 
instead  of the ion acoustic wave.   It  could  also be interesting to see if the  zero 
sound mode could be supported in the low mass brown  dwarfs stars studied in 
\cite {Berezhiani:2010db},  where densities are $\rho \sim 10^3~{\rm g/cm^3}$.
  
Finally, the  interaction between the electrons due to the exchange of the zero sound  mode
 (or the ion acoustic  wave)
would be strong if  the momentum transfer in the two-by-two 
electron scattering amplitude  is  near the pole of the zero sound  (or the ion acoustic  wave), i.e., is at $\omega\simeq xv_Fk$, with  
$x \gg 1$  (or with $x \ll 1$ for the ion acoustic wave). 
This interaction,  if attractive for a certain domain of momenta,  
might lead to formation of bound states, or loosely bound states such as Cooper pairs. 
While naive kinematical arguments suggest that the  momentum transfer  is not close to the 
zero sound pole as long as $x\gg 1$  (or with $x \ll 1$ for the ion acoustic wave),  
the issue needs careful study, presumably via  
the Schwinger - Dyson equation,  to see if such interactions can be used to 
trigger Cooper instability  and   produce a gap of  a physically  meaningful magnitude.

\vskip  1cm

\noindent {\bf Acknowledgments:}  The authors would like  to thank Paul Chaikin, Sergei Dubovsky,  Andy Millis, 
and Boris Spivak, for helpful  remarks.  Work of GG  is  supported by NASA grant NNX12AF86G S06, and 
NSF grant PHY-1316452. RAR is supported by DOE grant DE-SC0011941.  GG thanks the Physics Department of East China Norma University for warm hospitality.

\bibliographystyle{apsrev4-1}
\bibliography{ZeroSound}

\begin{thebibliography}{17}%
\makeatletter
\providecommand \@ifxundefined [1]{%
 \@ifx{#1\undefined}
}%
\providecommand \@ifnum [1]{%
 \ifnum #1\expandafter \@firstoftwo
 \else \expandafter \@secondoftwo
 \fi
}%
\providecommand \@ifx [1]{%
 \ifx #1\expandafter \@firstoftwo
 \else \expandafter \@secondoftwo
 \fi
}%
\providecommand \natexlab [1]{#1}%
\providecommand \enquote  [1]{``#1''}%
\providecommand \bibnamefont  [1]{#1}%
\providecommand \bibfnamefont [1]{#1}%
\providecommand \citenamefont [1]{#1}%
\providecommand \href@noop [0]{\@secondoftwo}%
\providecommand \href [0]{\begingroup \@sanitize@url \@href}%
\providecommand \@href[1]{\@@startlink{#1}\@@href}%
\providecommand \@@href[1]{\endgroup#1\@@endlink}%
\providecommand \@sanitize@url [0]{\catcode `\\12\catcode `\$12\catcode
  `\&12\catcode `\#12\catcode `\^12\catcode `\_12\catcode `\%12\relax}%
\providecommand \@@startlink[1]{}%
\providecommand \@@endlink[0]{}%
\providecommand \url  [0]{\begingroup\@sanitize@url \@url }%
\providecommand \@url [1]{\endgroup\@href {#1}{\urlprefix }}%
\providecommand \urlprefix  [0]{URL }%
\providecommand \Eprint [0]{\href }%
\providecommand \doibase [0]{http://dx.doi.org/}%
\providecommand \selectlanguage [0]{\@gobble}%
\providecommand \bibinfo  [0]{\@secondoftwo}%
\providecommand \bibfield  [0]{\@secondoftwo}%
\providecommand \translation [1]{[#1]}%
\providecommand \BibitemOpen [0]{}%
\providecommand \bibitemStop [0]{}%
\providecommand \bibitemNoStop [0]{.\EOS\space}%
\providecommand \EOS [0]{\spacefactor3000\relax}%
\providecommand \BibitemShut  [1]{\csname bibitem#1\endcsname}%
\let\auto@bib@innerbib\@empty
\bibitem [{\citenamefont {Shapiro}\ and\ \citenamefont
  {Teukolsky}(1983)}]{Teukolsky}%
  \BibitemOpen
  \bibfield  {author} {\bibinfo {author} {\bibfnamefont {S.~L.}\ \bibnamefont
  {Shapiro}}\ and\ \bibinfo {author} {\bibfnamefont {S.~A.}\ \bibnamefont
  {Teukolsky}},\ }\href@noop {} {\emph {\bibinfo {title} {Black holes, white
  dwarfs and neutron stars}}}\ (\bibinfo  {publisher} {John Wiley \& Sons},\
  \bibinfo {year} {1983})\BibitemShut {NoStop}%
\bibitem [{\citenamefont {Mestel}\ and\ \citenamefont
  {Ruderman}(1967)}]{MestelRuderman}%
  \BibitemOpen
  \bibfield  {author} {\bibinfo {author} {\bibfnamefont {L.}~\bibnamefont
  {Mestel}}\ and\ \bibinfo {author} {\bibfnamefont {M.}~\bibnamefont
  {Ruderman}},\ }\href@noop {} {\bibfield  {journal} {\bibinfo  {journal}
  {Monthly Notices of the Royal Astronomical Society}\ }\textbf {\bibinfo
  {volume} {136}},\ \bibinfo {pages} {27} (\bibinfo {year} {1967})}\BibitemShut
  {NoStop}%
\bibitem [{\citenamefont {Gabadadze}\ and\ \citenamefont
  {Rosen}(2008)}]{Gabadadze:2008mx}%
  \BibitemOpen
  \bibfield  {author} {\bibinfo {author} {\bibfnamefont {G.}~\bibnamefont
  {Gabadadze}}\ and\ \bibinfo {author} {\bibfnamefont {R.~A.}\ \bibnamefont
  {Rosen}},\ }\href {\doibase 10.1088/1475-7516/2008/10/030} {\bibfield
  {journal} {\bibinfo  {journal} {JCAP}\ }\textbf {\bibinfo {volume} {0810}},\
  \bibinfo {pages} {030} (\bibinfo {year} {2008})},\ \Eprint
  {http://arxiv.org/abs/0806.3692} {arXiv:0806.3692 [astro-ph]} \BibitemShut
  {NoStop}%
\bibitem [{\citenamefont {Ichimaru}(1982)}]{Ichimaru:1982zz}%
  \BibitemOpen
  \bibfield  {author} {\bibinfo {author} {\bibfnamefont {S.}~\bibnamefont
  {Ichimaru}},\ }\href {\doibase 10.1103/RevModPhys.54.1017} {\bibfield
  {journal} {\bibinfo  {journal} {Rev. Mod. Phys.}\ }\textbf {\bibinfo {volume}
  {54}},\ \bibinfo {pages} {1017} (\bibinfo {year} {1982})}\BibitemShut
  {NoStop}%
\bibitem [{\citenamefont {Gabadadze}\ and\ \citenamefont
  {Rosen}(2010)}]{Gabadadze:2009jb}%
  \BibitemOpen
  \bibfield  {author} {\bibinfo {author} {\bibfnamefont {G.}~\bibnamefont
  {Gabadadze}}\ and\ \bibinfo {author} {\bibfnamefont {R.~A.}\ \bibnamefont
  {Rosen}},\ }\href {\doibase 10.1088/1475-7516/2010/04/028} {\bibfield
  {journal} {\bibinfo  {journal} {JCAP}\ }\textbf {\bibinfo {volume} {1004}},\
  \bibinfo {pages} {028} (\bibinfo {year} {2010})},\ \Eprint
  {http://arxiv.org/abs/0912.5270} {arXiv:0912.5270 [hep-ph]} \BibitemShut
  {NoStop}%
\bibitem [{\citenamefont {Gabadadze}\ and\ \citenamefont
  {Rosen}(2009)}]{Gabadadze:2008pj}%
  \BibitemOpen
  \bibfield  {author} {\bibinfo {author} {\bibfnamefont {G.}~\bibnamefont
  {Gabadadze}}\ and\ \bibinfo {author} {\bibfnamefont {R.~A.}\ \bibnamefont
  {Rosen}},\ }\href {\doibase 10.1088/1475-7516/2009/02/016} {\bibfield
  {journal} {\bibinfo  {journal} {JCAP}\ }\textbf {\bibinfo {volume} {0902}},\
  \bibinfo {pages} {016} (\bibinfo {year} {2009})},\ \Eprint
  {http://arxiv.org/abs/0811.4423} {arXiv:0811.4423 [hep-th]} \BibitemShut
  {NoStop}%
\bibitem [{\citenamefont {Gabadadze}\ and\ \citenamefont
  {Pirtskhalava}(2009)}]{Gabadadze:2009dz}%
  \BibitemOpen
  \bibfield  {author} {\bibinfo {author} {\bibfnamefont {G.}~\bibnamefont
  {Gabadadze}}\ and\ \bibinfo {author} {\bibfnamefont {D.}~\bibnamefont
  {Pirtskhalava}},\ }\href {\doibase 10.1088/1475-7516/2009/05/017} {\bibfield
  {journal} {\bibinfo  {journal} {JCAP}\ }\textbf {\bibinfo {volume} {0905}},\
  \bibinfo {pages} {017} (\bibinfo {year} {2009})},\ \Eprint
  {http://arxiv.org/abs/0904.4267} {arXiv:0904.4267 [hep-th]} \BibitemShut
  {NoStop}%
\bibitem [{\citenamefont {Bedaque}\ \emph {et~al.}(2012)\citenamefont
  {Bedaque}, \citenamefont {Berkowitz},\ and\ \citenamefont
  {Cherman}}]{Bedaque:2011hs}%
  \BibitemOpen
  \bibfield  {author} {\bibinfo {author} {\bibfnamefont {P.~F.}\ \bibnamefont
  {Bedaque}}, \bibinfo {author} {\bibfnamefont {E.}~\bibnamefont {Berkowitz}},
  \ and\ \bibinfo {author} {\bibfnamefont {A.}~\bibnamefont {Cherman}},\ }\href
  {\doibase 10.1088/0004-637X/749/1/5} {\bibfield  {journal} {\bibinfo
  {journal} {Astrophys. J.}\ }\textbf {\bibinfo {volume} {749}},\ \bibinfo
  {pages} {5} (\bibinfo {year} {2012})},\ \Eprint
  {http://arxiv.org/abs/1111.1343} {arXiv:1111.1343 [nucl-th]} \BibitemShut
  {NoStop}%
\bibitem [{\citenamefont {Le~Bellac}(1996)}]{LeBellac:1996}%
  \BibitemOpen
  \bibfield  {author} {\bibinfo {author} {\bibfnamefont {M.}~\bibnamefont
  {Le~Bellac}},\ }\href@noop {} {\emph {\bibinfo {title} {{Thermal Field
  Theory}}}}\ (\bibinfo  {publisher} {Cambridge University Press},\ \bibinfo
  {address} {UK},\ \bibinfo {year} {1996})\BibitemShut {NoStop}%
\bibitem [{\citenamefont {Kapusta}\ and\ \citenamefont
  {Gale}(2006)}]{Kapusta:2006pm}%
  \BibitemOpen
  \bibfield  {author} {\bibinfo {author} {\bibfnamefont {J.}~\bibnamefont
  {Kapusta}}\ and\ \bibinfo {author} {\bibfnamefont {C.}~\bibnamefont {Gale}},\
  }\href@noop {} {\emph {\bibinfo {title} {{Finite-temperature field theory:
  Principles and applications}}}}\ (\bibinfo  {publisher} {Cambridge University
  Press},\ \bibinfo {address} {UK},\ \bibinfo {year} {2006})\BibitemShut
  {NoStop}%
\bibitem [{\citenamefont {Fetter}\ and\ \citenamefont
  {Walecka}(2003)}]{fetterwalecka}%
  \BibitemOpen
  \bibfield  {author} {\bibinfo {author} {\bibfnamefont {A.~L.}\ \bibnamefont
  {Fetter}}\ and\ \bibinfo {author} {\bibfnamefont {J.~D.}\ \bibnamefont
  {Walecka}},\ }\href@noop {} {\emph {\bibinfo {title} {Quantum Theory of
  Many-Particle Systems}}}\ (\bibinfo  {publisher} {Dover Publications},\
  \bibinfo {year} {2003})\BibitemShut {NoStop}%
\bibitem [{\citenamefont {Abrikosov}\ \emph {et~al.}(1975)\citenamefont
  {Abrikosov}, \citenamefont {Gorkov},\ and\ \citenamefont
  {Dzyaloshinski}}]{Dzyaloshinsky}%
  \BibitemOpen
  \bibfield  {author} {\bibinfo {author} {\bibfnamefont {A.~A.}\ \bibnamefont
  {Abrikosov}}, \bibinfo {author} {\bibfnamefont {L.~P.}\ \bibnamefont
  {Gorkov}}, \ and\ \bibinfo {author} {\bibfnamefont {I.~E.}\ \bibnamefont
  {Dzyaloshinski}},\ }\href@noop {} {\emph {\bibinfo {title} {Methods of
  quantum field theory in statistical physics}}}\ (\bibinfo  {publisher}
  {Courier Corporation},\ \bibinfo {year} {1975})\BibitemShut {NoStop}%
\bibitem [{\citenamefont {Berezhiani}\ \emph {et~al.}(2010)\citenamefont
  {Berezhiani}, \citenamefont {Gabadadze},\ and\ \citenamefont
  {Pirtskhalava}}]{Berezhiani:2010db}%
  \BibitemOpen
  \bibfield  {author} {\bibinfo {author} {\bibfnamefont {L.}~\bibnamefont
  {Berezhiani}}, \bibinfo {author} {\bibfnamefont {G.}~\bibnamefont
  {Gabadadze}}, \ and\ \bibinfo {author} {\bibfnamefont {D.}~\bibnamefont
  {Pirtskhalava}},\ }\href {\doibase 10.1007/JHEP04(2010)122} {\bibfield
  {journal} {\bibinfo  {journal} {JHEP}\ }\textbf {\bibinfo {volume} {04}},\
  \bibinfo {pages} {122} (\bibinfo {year} {2010})},\ \Eprint
  {http://arxiv.org/abs/1003.0865} {arXiv:1003.0865 [hep-ph]} \BibitemShut
  {NoStop}%
\bibitem [{\citenamefont {Dolgov}\ \emph
  {et~al.}(2009{\natexlab{a}})\citenamefont {Dolgov}, \citenamefont {Lepidi},\
  and\ \citenamefont {Piccinelli}}]{Dolgov:2008pe}%
  \BibitemOpen
  \bibfield  {author} {\bibinfo {author} {\bibfnamefont {A.~D.}\ \bibnamefont
  {Dolgov}}, \bibinfo {author} {\bibfnamefont {A.}~\bibnamefont {Lepidi}}, \
  and\ \bibinfo {author} {\bibfnamefont {G.}~\bibnamefont {Piccinelli}},\
  }\href {\doibase 10.1088/1475-7516/2009/02/027} {\bibfield  {journal}
  {\bibinfo  {journal} {JCAP}\ }\textbf {\bibinfo {volume} {0902}},\ \bibinfo
  {pages} {027} (\bibinfo {year} {2009}{\natexlab{a}})},\ \Eprint
  {http://arxiv.org/abs/0811.4406} {arXiv:0811.4406 [hep-th]} \BibitemShut
  {NoStop}%
\bibitem [{\citenamefont {Dolgov}\ \emph
  {et~al.}(2009{\natexlab{b}})\citenamefont {Dolgov}, \citenamefont {Lepidi},\
  and\ \citenamefont {Piccinelli}}]{Dolgov:2009yt}%
  \BibitemOpen
  \bibfield  {author} {\bibinfo {author} {\bibfnamefont {A.~D.}\ \bibnamefont
  {Dolgov}}, \bibinfo {author} {\bibfnamefont {A.}~\bibnamefont {Lepidi}}, \
  and\ \bibinfo {author} {\bibfnamefont {G.}~\bibnamefont {Piccinelli}},\
  }\href {\doibase 10.1103/PhysRevD.80.125009} {\bibfield  {journal} {\bibinfo
  {journal} {Phys.Rev.}\ }\textbf {\bibinfo {volume} {D80}},\ \bibinfo {pages}
  {125009} (\bibinfo {year} {2009}{\natexlab{b}})},\ \Eprint
  {http://arxiv.org/abs/0905.4422} {arXiv:0905.4422 [hep-ph]} \BibitemShut
  {NoStop}%
\bibitem [{\citenamefont {D'Antona}\ and\ \citenamefont
  {Mazzitelli}(1990)}]{DAntona:1990ea}%
  \BibitemOpen
  \bibfield  {author} {\bibinfo {author} {\bibfnamefont {F.}~\bibnamefont
  {D'Antona}}\ and\ \bibinfo {author} {\bibfnamefont {I.}~\bibnamefont
  {Mazzitelli}},\ }\href {\doibase 10.1146/annurev.aa.28.090190.001035}
  {\bibfield  {journal} {\bibinfo  {journal} {Ann. Rev. Astron. Astrophys.}\
  }\textbf {\bibinfo {volume} {28}},\ \bibinfo {pages} {139} (\bibinfo {year}
  {1990})}\BibitemShut {NoStop}%
\bibitem [{\citenamefont {Hansen}\ and\ \citenamefont
  {Liebert}(2003)}]{Hansen:2003bc}%
  \BibitemOpen
  \bibfield  {author} {\bibinfo {author} {\bibfnamefont {B.~M.}\ \bibnamefont
  {Hansen}}\ and\ \bibinfo {author} {\bibfnamefont {J.}~\bibnamefont
  {Liebert}},\ }\href {\doibase 10.1146/annurev.astro.41.081401.155117}
  {\bibfield  {journal} {\bibinfo  {journal} {Ann. Rev. Astron. Astrophys.}\
  }\textbf {\bibinfo {volume} {41}},\ \bibinfo {pages} {465} (\bibinfo {year}
  {2003})}\BibitemShut {NoStop}%
\end{thebibliography}%

\end{document}